\documentclass[11pt]{article}
\usepackage[top=3cm,bottom=2.8cm,left=2.6cm,right=2.6cm]{geometry}

\usepackage{hyperref}
\hypersetup{colorlinks,citecolor=black,filecolor=black,linkcolor=black,
urlcolor=black,bookmarksnumbered}

\usepackage{amsmath,amsfonts,amssymb,amsthm,amscd,bbm,mathrsfs}
\usepackage{cancel}
\usepackage{listings}
\usepackage{moreverb}
\usepackage{booktabs}
\usepackage{epsfig}
\usepackage{cite}
\usepackage{algorithm}
\usepackage[]{algpseudocode}

\providecommand{\keywords}[1]{\noindent{\textbf{\textit{Keywords---}}} #1}
\providecommand{\norm}[1]{\lVert #1\rVert }

\begin{document}

\title{Synthesis Method for the Spherical 4R Mechanism with Minimum 
Center of Mass Acceleration}

\author{{\small
$^a$O. Mendoza-Trejo\footnote{omendoza@cinvestav.mx},~
$^a$Carlos A. Cruz-Villar\footnote{cacruz@cinvestav.mx},~ 
$^b$R. Pe\'on-Escalante\footnote{rpeon@uady.mx},~
$^b$M. A. Zambrano-Arjona\footnote{miguel.zambrano@uady.mx},~
$^b$F. Pe\~nu\~nuri\footnote{francisco.pa@uady.mx}}\\
{\footnotesize \textit{$^a$Cinvestav-IPN, Departamento de Ingenier\'ia 
El\'ectrica, Av. IPN 2508, A. P. 14-740, 07300, M\'e\-xi\-co D.F., 
M\'exico.}}\\
{\footnotesize \textit{$^b$Facultad de Ingenier\'ia, Universidad 
Aut\'onoma de Yucat\'an, A.P. 150, Cordemex, M\'erida, Yucat\'an, 
M\'exico.}}
}
\date{}
\maketitle
\begin{abstract}
In the mechanisms area, minimization of the magnitude of the 
acceleration of the center of mass (ACoM) implies shaking force 
balancing.  This article shows an efficient optimum 
synthesis method for minimum acceleration of the center of mass of a 
spherical 4R mechanism by using dual functions, as well as, the 
counterweights balancing method. Once the dual function for ACoM has 
been formulated, one can minimize the shaking forces from a kinematic 
point of view. We present the synthesis of a spherical 4R mechanism for 
the case of a path generation task. The synthesis process involves the 
optimization of two objective functions, this multiobjective problem is 
solved by using the weighted sum method implemented in the evolutionary 
algorithm known as Differential Evolution. Our results shows that the
magnitude of the ACoM can be reduced about 20 times. Which for the 
presented synthesis, is traduced in a reduction of about 89\% for the 
shacking forces.

\vspace*{0.25cm}
\end{abstract}%

\section{Introduction}
A mechanism is said to be dynamically balanced if there is 
an elimination of dynamic reaction forces  and torques (moments) on the 
base \cite{Gosselin2004,Qimi2010,Wijk2009,Wijk2013}.
The use of counterweights \cite{Bagci1982,Feng1990,Smith1999,
Schutter2006} and springs \cite{Gosselin1999, Agrawal2005, Deepak2012} 
are the two most common methods for balancing a mechanism.

The balancing of mechanisms is a very active field in the mechanisms 
area, and there is a big amount of bibliography on such a topic, we 
refer the reader to works \cite{Lowen1983, Arakelian2005}  where the 
balancing of mechanisms is reviewed.  Specifically regarding to the 
concept of the center of mass, it is worthwhile to mention 
\cite{Rosario2005, Volkert2012, Arakelian2012}. In \cite{Rosario2005},  
the balancing condition of parallel robots is obtained by expressing the 
position vector of the  center of mass  as a function of the 
position and orientation of the platform, and of six actuated prismatic 
joints. In \cite{Volkert2012}, a  synthesis method for linkages with the 
center of mass at an invariant link point is presented. 
In \cite{Arakelian2012}, the shaking force minimization of high-speed 
robots is done via optimal control of the acceleration of the total mass 
center of moving links.

The problem of cancellation or minimization of the total external force 
can be addressed from a pure kinematic point of view by minimizing the 
ACoM. The idea of an optimization approach has 
been successfully applied in \cite{Pennestri1991,Li1992,Himanshu2007,
Subir2008,Saravanana2008}. The smaller the ACoM, the smaller the
shaking force in the mechanism. 
In the present work we show how, by using dual numbers \cite{Clif1873, 
Brodsky1999, Ettore2008} and the counterweights balancing method, a 
simple and computationally efficient optimum synthesis method with 
minimum center of mass acceleration can be achieved. First, we provide a 
method where the acceleration of the center of mass can be precisely and 
efficiently calculated. Second, since in some cases, a mass 
redistribution can become equal to zero the acceleration of the center 
of mass, we use an optimization based methodology, to find such a mass 
redistribution.

Our study is focused on the path generation task for a spherical 4R 
mechanism.  The problem involves two objective functions to be 
optimized, namely the objective function to fulfill the desired 
trajectory and the objective function related to the acceleration of the 
center of mass. This multiobjective problem is solved by using the 
weighted sum method, which is implemented on the evolutionary algorithm 
known as differential evolution (DE) \cite{Price_Storn2005}. The 
synthesized mechanism was constructed, and actuated by a dc motor 
controlled by a classical PID law. This allow us to compare the required 
power (and energy used) to operate the balanced and unbalanced 
mechanisms. The experimental measurement of the energy consumption was 
contrasted with its theoretical calculation which is conducted by using 
the work--energy theorem, providing in this way an experimental 
validation for the reduction of the ACoM in the balanced mechanism.

The rest of the paper is organized as follows, section \ref{sec2} 
presents the essentials of our method. There, the balancing method, the 
dual number approach to obtain derivatives and the optimization method 
are presented. In section \ref{sec3} a Spherical 4R mechanism is 
synthesized for a prescribed path generation task with minimum center of 
mass acceleration. This section shows all the parameters and formulas 
that need to be written in its dual form in order to obtain the 
acceleration of the center of mass vector. Also a theoretical calculation
of the energy consumption is presented. Section \ref{Results} shows 
the results along some discussions. Finally, the conclusions are 
presented in section \ref{conclu}. 

\section{Synthesis Method}\label{sec2}
In general, a mechanism has an arbitrary time-varying ACoM. So our aim 
is to minimize the magnitude of the ACoM. For that purpose we need a 
precise and efficient method to obtain first and second derivatives, as 
well as an optimization method that allow us to cancel or minimize the 
ACoM. This section shows the implementation of the dual numbers to 
obtain derivatives and the method used for balancing.

\subsection{Minimum Acceleration of Center of Mass and Counterweight 
Method }\label{subsec2.1}
There are several methods for balancing mechanisms, each one having 
advantages and disadvantages \cite{Lowen1983, Arakelian2005, Wijk2009}. 
Since a mathematical point of view, more free parameters are introduced 
to the synthesis problem to balance a mechanism. We implement this 
approach for balancing a spherical 4R (Fig. \ref{figure1}) mechanism by 
adding counterweights \cite{Bagci1982,Feng1990,Smith1999,Rosario2005} to 
the links as shown in Fig. \ref{figure2}, therefore we have more 
independent variables in order to minimize the ACoM of the Spherical 4R 
mechanism. 

\subsection{Dual numbers approach}\label{subsec2.2}
Usually, the process of calculating a derivative is not difficult. 
However, for the case of a spherical mechanism, obtaining first and 
second derivatives of the position vector for the center of mass is not 
simple. Even when such derivatives can be explicitly obtained, the 
resulting expressions could be of great complexity and useless for 
practical purposes. An alternative solution is to numerically calculate 
such derivatives. Nevertheless, traditional methods for calculating 
numerical derivatives (finite-differences) are not efficient enough to 
be used in the optimum synthesis of mechanisms. Moreover, they are 
subject to both truncation and subtractive cancellation errors. This is 
an issue to be dealt with in the optimum synthesis of mechanisms, since 
in some cases the errors introduced by the numerical derivative method 
are of the same order than the quantity to be optimized.

A different approach that is not subject to the above mentioned errors 
can be constructed by using  dual numbers \cite{Gu1987,Cheng1994,
Penunuri2013}. In order to make this paper self contained we briefly 
review the essential ideas following \cite{Penunuri2013} and bearing in 
mind a numerical implementation. 

\subsubsection*{Dual numbers and derivatives}
A dual number $\hat{r}$ is a number of the form 
\begin{equation}
\hat{r}= a +\epsilon \,b,
\end{equation}
where $a$ (the real coefficient) and $b$ (the dual coefficient) are real 
numbers and $\epsilon^2=0$. As in the case of complex numbers, a dual 
number can also be defined as ordered pairs
\begin{equation}\label{xDual}
\hat{r}=[a,b].
\end{equation}
 
The algebraic rules for dual numbers can be found elsewhere in the 
literature, see for example  \cite{Clif1873, Brodsky1999,Ettore2008}.

For a function $f:\mathbb{R}\to \mathbb{R}$ which accepts a Taylor 
expansion we have:
\begin{equation} \label{TayS}
f(x+h)=f(x) \, + \, f'(x) h \, + \, \frac{f''(x)}{2} h^2 \,+\, \dots \,
+\, O(h^3).
\end{equation}

Now, considering the particular dual number $\hat{x}=x+\epsilon$, i.e., 
a dual number where the coefficient of the nilpotent $\epsilon$ is equal 
to one we have $ \hat{f}(\hat{x}) = f(x) + f'(x) \epsilon.$ In the 
notation of (\ref{xDual}) we may write
\begin{equation} \label{dfv}
\hat{f}(\hat{x}) = \left[ f(x),f'(x) \right].
\end{equation} 
So, if instead of calculating over the reals, we calculate over the 
\textit{duals}, we end up with a dual function where the real part is 
the function itself and the dual part its derivative. 

The extension to obtain the second derivative is straightforward. For 
this purpose we promote the $f'(x)$ function to be a 
dual function, that is:
\begin{equation}
 \hat{f}'(\hat{x})= \left[f'(x),f''(x) \right].
\end{equation}

The information of the function itself and its first and second 
derivative can be stored in a vector of three components.  We will use 
the notation 
\begin{equation}
\tilde{f}(x)=\left[f(x),f'(x),f''(x) \right ]
\end{equation}
to represent the extended dual version of the original function $f$.

The following pseudo-code implements the extended dual 
version $\tilde{f} =[f_0,f_1,f_2]$ of the function $f$, as a function of 
the dual variable $\tilde{g} = [g_0,g_1,g_2]$:

\begin{verbatim}
start the definition of the dual function 
declare the type of variables (if necessary)
dualf = [f_0(g_0),f_1(g_0)*g_1, f_2(g_0)*g_1^2 + f_1(g_0)*g_2 ]
end the definition of the dual function. 
\end{verbatim}
 
Following the above discussion, one can obtain the extended dual version 
of the center of mass vector from where its velocity and acceleration 
are  directly obtained. The derivatives are calculated to the same 
precision of the implemented standard functions of the used programming 
language, and its calculation is reduced to a simple function evaluation 
without the need of a limit calculation process. As an added value to 
our work, some useful functions common in the rotational kinematics are 
dualized for the first time and provided as additional material to this 
article\footnote{A file ({\tt Dual\_ACoM\_files.zip}) containing the 
code of such functions can be downloaded from 
\url{http://www.meca.cinvestav.mx/personal/cacruz/archivos-ccv/} }. 
So they will be of great help to any one interested in applying the dual 
numbers to obtain derivatives. Since we are interested in the optimal 
synthesis of mechanisms (fast and efficient programs are desired), we 
have coded the dual functions in Fortran but a translation to a more 
user-friendly language is almost direct. 

\subsection{Optimization method}
The optimal dimensional synthesis is conducted by using the evolutionary 
DE method \cite{Price_Storn2005}. This method has been successfully 
applied to the optimal dimensional synthesis of mechanisms 
\cite{Cabrera2007,Bulatovic2009,NewVillage2011,Cabrera2011} 
and mainly consists of three operators:  mutation, crossover and 
selection; applied to a possible solution or individual $\mathbf{x}$.  
Below an outline of the DE method, following  \cite{Price_Storn2005} is 
presented.

\subsubsection{Differential Evolution Algorithm}
\begin{enumerate}
\item 
The population is described by:
\begin{eqnarray}\nonumber \label{ClassicDE}
\mathbf{P_{\mathbf{x},g}} &=&(\mathbf{x}_{i,g}), ~i=1,...,m;~~g=0,...,
g_{\text{max}}\\
\mathbf{x}_{i;g}&=&(x^j_{i;g}), ~j=1,...,D;
\end{eqnarray}
where $D$ represents the dimension of $\mathbf{x}$, $m$ represents the 
number of individuals, and $g$ is the generation.

\item
Initialization of population.
\begin{equation*}
x^j_{i;0}=rand^j(0,1)\cdot (b^j_{\text{U}}-b^j_{\text{L}})+
b^j_{\text{L}}.
\end{equation*}
Vectors $\mathbf{b}_{\text{U}}$ and $\mathbf{b}_{\text{L}}$ are the 
limits of parameters.
$rand^j(0,1)$ represents uniformly distributed random number in $[0,1)$.
Superscript $j$ means that a random value for each parameter is 
generated.
\item
Mutation.
\begin{equation}
\mathbf{v}_{i;g}=\mathbf{x}_{r_0;g}+F\cdot (\mathbf{x}_{r_1;g}-
\mathbf{x}_{r_2;g}).
\end{equation}
The main difference between DE and other evolutionary algorithms like 
genetic algorithms is due to this mutation operator. 
$\mathbf{x}_{r_0;g}$ is called the base vector which is perturbed by the 
difference of other two vectors.
 $r_0, ~r_1,~r_2~\in \{1,2,...m\},~ r_1\neq r_2\neq r_3\neq i$.
$F$, the mutation scale factor, is a real number greater than zero.

\item
Crossover.\\
It uses a dual recombination of vectors to generate the trial vector:
\begin{equation}\label{DECrossover}
\mathbf{u}_{i;g} = u^j_{i;g}=
\begin{cases}
v^j_{i;g} & \text{if $rand^j(0,1)\leqslant Cr$ or $j=j_{\text{rand}}$}
\vspace*{0.3cm}\\
x^j_{i;g} & \text{otherwise.}
\end{cases}
\end{equation}
The crossover probability, $Cr \in [0,1]$, is a user-defined value, 
$j_{\text{rand}} \in [1,D]$.
\item
Selection.\\
The selection is made according to
\begin{equation}\label{finClassicDE}
\mathbf{x}_{i;g+1}=\left \lbrace \begin{array}{l} \mathbf{u}_{i;g}~~
\text{if $f(\mathbf{u}_{i;g})\leqslant f(\mathbf{x}_{i;g})$}  \\
\mathbf{x}_{i;g}~~\text{otherwise}
 \end{array}\right. 
 \end{equation}
\end{enumerate}

\subsubsection{Objective function for the prescribed path generation 
task}
In this study we are interested in the prescribed path generation task. 
So, besides an objective function related to the minimization of the 
ACoM, we also consider an objective function intended to fulfill the 
path generation task.

In the path generation problem,  $n$ points are given and a mechanism 
passing through these desired points is required. In order to synthesize 
such a mechanism (in this study a spherical 4R mechanism), we minimize 
the mean structural error (root mean square error) which  is defined as:
\begin{equation}\label{objectivef}
f_{path}(\theta_1,\phi_1,\eta_1,\psi,\beta,\gamma,\alpha_1,\alpha_2,
\alpha_3,\alpha_4)=\left(\frac{1}{n}\sum_{i=1}^n \norm{{\mathbf r}_{di}-
{\mathbf r}_{\text{gen}i}}^2\right)^{1/2},
\end{equation}
where ${\mathbf r}_{di}-{\mathbf r}_{\text{gen}i}$ is the difference 
between the $i$-th desired point ${\mathbf r}_{di}$ and   the $i$-th 
generated point ${\mathbf r}_{\text{gen}i}$.
The $\theta_1$ parameter, is the angle  for the input link corresponding 
to the first point. In the prescribed path generation task the angle 
corresponding to the other points are given. 

\subsubsection{Objective function for minimum center of  mass 
acceleration}
In order to obtain a mechanism with minimum ACoM we minimize 
the objective function
  \begin{equation}\label{fcm}
  f_{cm}(\theta_1,\phi_1,\eta_1,\psi,\beta,\gamma,\alpha_1,\alpha_2,
  \alpha_3,\alpha_4,e_1,e_2,e_3,e_4)=\left(\frac{1}{n}\sum_{i=1}^n 
  \norm{\mathbf{\ddot{r}}_{cm;i}}^2\right)^{1/2}
  \end{equation}
  where $\mathbf{\ddot{r}}_{cm;i}$ is the acceleration of the center of 
  mass of the mechanism when it is passing through the $i$-th point. 

\section{Synthesis of a Spherical 4R Mechanism}\label{sec3}
This section presents the application of the method above explained, to 
the synthesis of a spherical 4R mechanism for a path generation task, 
with minimum ACoM. Table \ref{table64p} (as presented in 
\cite{Mullineux2011}) shows the desired points. This problem was
originally studied in \cite{ChuSun2010} but there, the desired points 
are not on a unit sphere. 
\begin{figure}[htb]
\begin{center}
\includegraphics[scale=0.4]{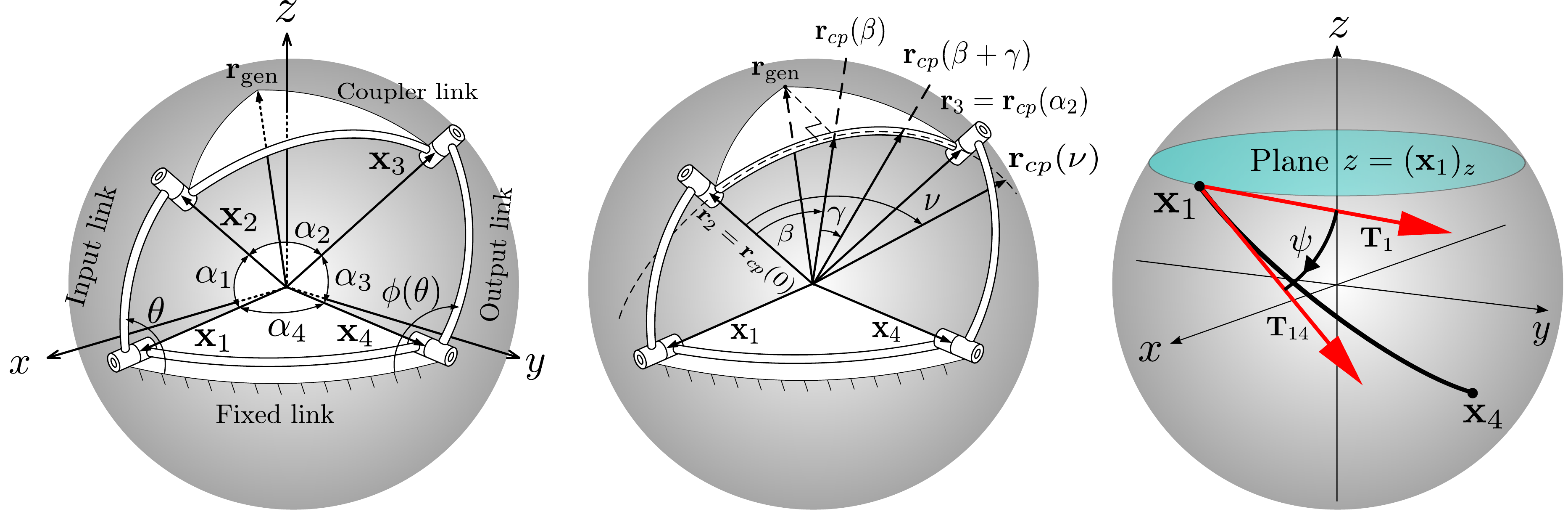} 
\caption{Spherical 4R Mechanism and variables used to obtain 
$\mathbf{r}_{\text{gen}}$.}
\label{figure1}
\end{center}
\end{figure}
\begin{table}[ht]
\caption{Desired points for the prescribed timing path generation task.}
\vspace*{0.1cm}
\centering
\scalebox{0.8}{
\begin{tabular}{c c c c }
\toprule Point number & Point & Point number & Point\\
\midrule 1 & (0.85737, -0.18481, 0.48037) & 33 & (0.7887, -0.60370, 0.11578) \\
               2 & (0.82985, -0.20167, 0.52030) & 34 & (0.80152, -0.59270, 0.07900) \\
               3 & (0.80241, -0.21996, 0.55478) & 35 & (0.81378, -0.57959, 0.04311) \\
               4 & (0.77567, -0.23967, 0.58389) & 36 & (0.82552, -0.56433, 0.00841) \\
               5 & (0.75011, -0.26056, 0.60785) & 37 & (0.83678, -0.54700, -0.02478) \\
               6 & (0.72607, -0.28244, 0.62693) & 38 & (0.84759, -0.52763, -0.05611) \\
               7 & (0.70381, -0.30515, 0.64152) & 39 & (0.85807, -0.50641, -0.08530) \\
               8 & (0.68352, -0.32833, 0.65193) & 40 & (0.86819, -0.48344, -0.11200) \\
               9 & (0.66533, -0.35185, 0.65844) & 41 & (0.87804, -0.45889, -0.13596) \\
             10 & (0.64933, -0.37537, 0.66144) & 42 & (0.88763, -0.43304, -0.15689) \\
             11 & (0.63559, -0.39867, 0.66115) & 43 & (0.89704, -0.40611, -0.17448) \\ 
             12 & (0.62415, -0.42159, 0.65781) & 44 & (0.90626, -0.37837, -0.18848) \\
             13 & (0.61504, -0.44389, 0.65167) & 45 & (0.91537, -0.35022, -0.19867) \\ 
             14 & (0.60833, -0.46541, 0.64293) & 46 & (0.92433, -0.32193, -0.20481) \\
             15 & (0.60396, -0.48596, 0.63170) & 47 & (0.93322, -0.29396, -0.20667) \\
             16 & (0.60196, -0.50548, 0.61819) & 48 & (0.94196, -0.26667, -0.20400) \\
             17 & (0.60230, -0.52381, 0.60241) & 49 & (0.95052, -0.24048, -0.19667) \\
             18 & (0.60485, -0.54085, 0.58448) & 50 & (0.95885, -0.21581, -0.18441) \\ 
             19 & (0.60959, -0.55656, 0.56448) & 51 & (0.96685, -0.19315, -0.16707) \\
             20 & (0.61637, -0.57081, 0.54244) & 52 & (0.97430, -0.17289, -0.14452) \\
             21 & (0.62500, -0.58363, 0.51844) & 52 & (0.98096, -0.15541, -0.11656) \\
             22 & (0.63530, -0.59489, 0.49248) & 54 & (0.98652, -0.14104, -0.08315) \\
             23 & (0.64700, -0.60456, 0.46467) & 55 & (0.99052, -0.13007, -0.04430) \\
             24 & (0.65989, -0.61259, 0.43507) & 56 & (0.99244, -0.12263, -0.00019) \\
             25 & (0.67370, -0.61896, 0.40378) & 57 & (0.99174, -0.11870, 0.04874) \\
             26 & (0.68811, -0.62363, 0.37093) & 58 & (0.98774, -0.11822, 0.10185) \\
             27 & (0.70293, -0.62652, 0.33674) & 59 & (0.98000, -0.12085, 0.15807) \\
             28 & (0.71789, -0.62759, 0.30133) & 60 & (0.96819, -0.12626, 0.21604) \\
             29 & (0.73274, -0.62678, 0.26500) & 61 & (0.95226, -0.13415, 0.27426) \\
             30 & (0.74737, -0.62404, 0.22800) & 62 & (0.93252, -0.14411, 0.33115) \\
             31 & (0.76167, -0.61933, 0.19059) & 63 & (0.90956, -0.15600, 0.38515) \\
             32 & (0.77544, -0.61256, 0.15307) & 64 & (0.88422, -0.16959, 0.43519) \\
\bottomrule
\end{tabular}
}
\label{table64p}
\end{table}

\subsection{Kinematics}
A spherical four-bar mechanism (Fig. \ref{figure1}) is a closed chain, 
consisting of four links and four revolute joints. These linkages have 
the property that every link in the system rotates about the same fixed 
point \cite{McCarthy2000}. Thus, in a spherical mechanism, any point in 
a moving body is confined to move within a spherical surface, and all 
spherical surfaces of motion are concentric. In order to find the 
coupler point curve $\mathbf{r}_{\text{gen}}$, we will use the 
parameters shown in Fig.  \ref{figure1}. Such equation has been already 
obtained in \cite{Penunuri2012}, however, we want to change to a more 
concise description by using the lengths of the links as parameters 
instead of the coordinates for the joints. Thus, the coupler-point curve 
can be written as: 
\begin{equation}\label{rgenx}
{\mathbf r}_{\text{gen}}(\theta,\phi_1,\eta_1,\psi,\beta,\gamma,
\alpha_1,\alpha_2,\alpha_3,\alpha_4)={\mathbf R}\left(\pi /2,
{\mathbf r}_{cp}(\beta)\right)\, {\mathbf r}_{cp}(\beta +\gamma),
\end{equation}
where $\phi_1$ and $\eta_1$ are the azimuth and polar angle for the 
point $\mathbf{x}_1$, ${\mathbf R}(\nu,\hat{\mathbf{w}})$ is the active 
rotation matrix of angle $\nu$ in direction of the unit vector 
$\hat{\mathbf{w}}$.
\begin{equation}\label{couplerpoint}
{\mathbf r}_{cp}(\nu) ={\mathbf R}(\nu,\hat{{\mathbf n}}_{23})\,
{\mathbf r}_2(\theta,\psi,\phi_1,\eta_1,\alpha_1,\alpha_4),
\end{equation}
\begin{equation}\label{n23}
\hat{\bf{n}}_{23}=\dfrac{{\mathbf r}_2(\theta,\psi,\phi_1,\eta_1,
\alpha_1,\alpha_4) \times {\mathbf r}_3 (\phi(\theta),\psi,\phi_1,
\eta_1,\alpha_1,\alpha_2,\alpha_3,\alpha_4)} 
{\norm{{\mathbf r}_2(\theta,\psi,\phi_1,\eta_1,\alpha_1,\alpha_4) \times 
{\mathbf r}_3 (\phi(\theta),\psi,\phi_1,\eta_1,\alpha_1,\alpha_2,
\alpha_3,\alpha_4)}},
\end{equation}
\begin{equation}\label{r2}
 {\mathbf r}_2(\theta,\psi,\phi_1,\eta_1,\alpha_1,\alpha_4) ={\mathbf R}
 (\theta,{\mathbf x}_1)\,\mathbf{R}(\alpha_1,\hat{\bf{n}}_{14})\,
 \mathbf{x}_1.
\end{equation}
For $\eta_1\neq 0,~ \pi$, we obtain\footnote{Similar formulas can be 
obtained when $\eta_1 = 0$ or $\eta_1 = \pi$ although a simple solution 
is to chose another basis, for instance $\lbrace\hat{\mathbf{k}},~
\hat{\mathbf{i}},~\hat{\mathbf{j}}\rbrace$.}
\begin{equation}\label{n14}
\hat{\bf{n}}_{14} = \frac{\mathbf{x}_1\times \mathbf{T}_{14}}{\norm{
\mathbf{x}_1\times \mathbf{T}_{14}}}
\end{equation}
\begin{equation}
\mathbf{T}_{14} = \mathbf{R}(\psi,\mathbf{x}_1)\,\mathbf{T}_1
\end{equation}
\begin{equation}
\mathbf{T}_1 = \frac{\partial  }{\partial \, u}\left(\mathbf{R}(u,
\hat{\mathbf{k}})\,\mathbf{x}_1\right)\bigg|_{u=0}
\end{equation}
 \begin{equation}\label{r3}
 {\mathbf r}_3(\phi(\theta),\psi,\phi_1,\eta_1,\alpha_1,\alpha_2,
 \alpha_3,\alpha_4)={\mathbf R}(\phi(\theta),-{\mathbf x}_4)\,\mathbf{R}
 (\alpha_3,-\hat{\mathbf{n}}_{14})\,\mathbf {x}_4,
\end{equation} 
\begin{equation}
\mathbf{x}_4 = \mathbf{R}(\alpha_4,\hat{\bf{n}}_{14})\,\mathbf{x}_1.
\end{equation}
Finally, the output angle $\phi(\theta)$ is given by:
\begin{align}\label{phiangleB}
\phi(\theta)=\, &2 \tan^{-1}\left(\frac{-A\pm \sqrt{A^2+B^2-C^2}}{C-B}
\right)\text{where}\\
A=\,&\sin \alpha _1 \sin \alpha _3 \sin \theta \nonumber\\
B=\,&\cos \alpha _1 \sin \alpha _3 \sin \alpha _4 -
\sin \text{$\alpha _1$} \sin \text{$\alpha _3$} \cos
\text{$\alpha _4$} \cos \theta \nonumber\\
C=\,&\cos \text{$\alpha _1$} \cos \text{$\alpha _3$}
\cos \text{$\alpha _4$} + \sin \text{$\alpha _1$} \cos 
\text{$\alpha _3 $} \sin \text{$\alpha _4$} \cos \theta -\cos 
\text{$\alpha _2 $}\nonumber.
\end{align}
It is worthwhile to mention that, since points 
$\mathbf{x}_2$ and $\mathbf{x}_3$ are not given as input parameters, we 
arbitrarily can choose any of the two branches. Notice however, that if 
points $\mathbf{x}_2$ and $\mathbf{x}_3$ are given as input parameters, 
then a careful choice of the branch is needed (see for example appendix 
C of \cite{Cervantes2009} for details).

\subsection{Center of Mass Acceleration}
\begin{figure}[htb]
\begin{center}
 \includegraphics[scale=0.27]{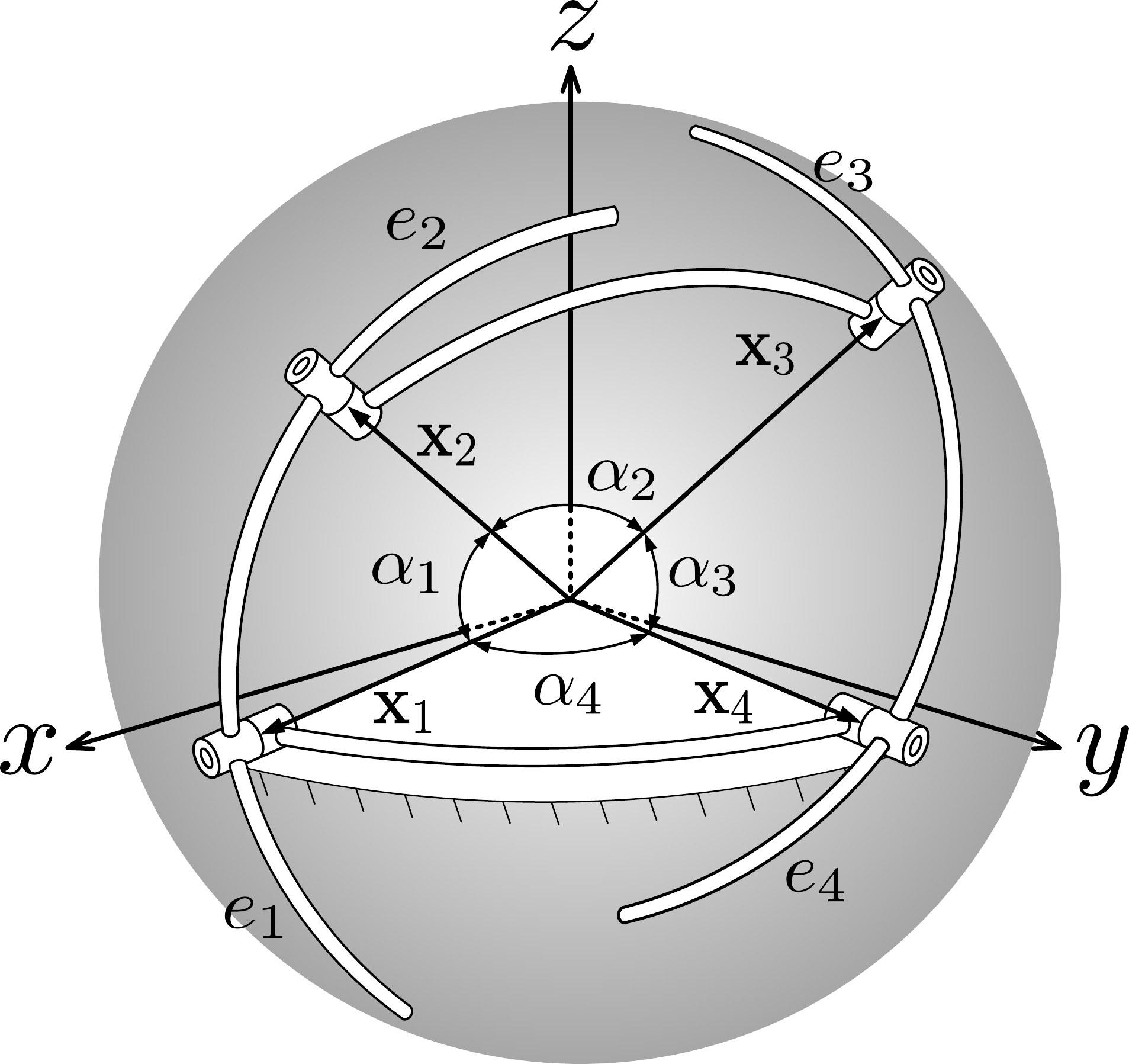} 
\caption{Extensions of the Input and Output Links.}
\label{figure2}
\end{center}
\end{figure}
Let us consider that a link has the geometry of a geodesic joining two 
points, $\mathbf{p}_1$ and $\mathbf{p}_2$ on the unit sphere, whose 
associated vectors subtends an arc $\alpha$. Assuming that the link is 
made of an homogeneous material, the center of mass vector of the link 
is
\begin{equation}
 \mathbf{r}_{cm;l}=\frac{2}{\alpha} \sin(\alpha/2) \,\mathbf{R}\,
 (\alpha /2, \hat{\mathbf{p}}_{12} )\,\mathbf{p}_1 ,
\end{equation}
where $ \hat{\mathbf{p}}_{12} =\mathbf{p}_1 \times \mathbf{p}_2/\Vert 
\mathbf{p}_1 \times \mathbf{p}_2  \Vert$.

Since the density of mass is a constant, the center of mass vector of 
the input link (without extensions) can be calculated as:
\begin{equation}
\mathbf{r}_{cm;l_1} = \frac{2}{\alpha_1}\,\sin(\alpha_1/2) \,
\mathbf{R}\,(\alpha_1 /2, \hat{\mathbf{n}}_{12} )\, \mathbf{x}_1 
\end{equation}
with 
\begin{equation}\label{n12}
\hat{\bf{n}}_{12}=\dfrac{\mathbf{x}_1 \times 
{\mathbf r}_2 (\theta,\psi,\phi_1,\eta_1,\alpha_1,\alpha_4)} 
{\norm{\mathbf{x}_1 \times {\mathbf r}_2(\theta,\psi,\phi_1,\eta_1,
\alpha_1,\alpha_4)}}.
\end{equation}

The center of mass for the other links and extensions (Fig. 
\ref{figure2}) can be similarly calculated. Once all the center of mass 
for the links and extensions have been calculated, the center of mass of 
the complete mechanism is:
\begin{equation}\label{rcmeq}
\mathbf{r}_{cm} = \sum _k s_k  \, \mathbf{r}_{cm;k}\Big{ /}\sum_ k s_k,
\end{equation}
where $s_k$ is the length of the $k-th$ link (or extension) for $k=e_1,
e_2,e_3,e_4,l_1,$ etc. 

Now, the acceleration of the center of mass will be
\begin{equation}\label{rcmdt2}
\mathbf{\ddot{r}}_{cm} = \ddot{\theta}\,\frac{\partial \mathbf{r}_{cm}}
{\partial \theta}\, + \,\dot{\theta}^2\,\frac{\partial^2\mathbf{r}_{cm}}
{\partial \theta^2}.
\end{equation}
Since it is possible to control the angular velocity of the input link, 
the calculation of the acceleration of the center of mass is reduced to 
calculate $\partial \mathbf{r}_{cm}/\partial \theta$ and 
$\partial^2\mathbf{r}_{cm}/\partial \theta^2$. Both first and second 
derivatives are obtained when we write Eq. (\ref{rcmeq}) in its dual 
form (see the \verb+rcm_dual_mod.f90+ file of the additional material).
In order to clarify, let us find the dual version for the center of mass
of the input link. Let $\widetilde{\mathbf{R}}(\tilde{\theta},
\tilde{\mathbf{u}})$ be the dual version of the
active rotation matrix of dual angle $\tilde{\theta}$ about the unit 
dual vector $\tilde{\mathbf{u}}$.
Then
\begin{equation}
 \tilde{\mathbf{r}}_{cm;l_1} = \frac{2}{\alpha_1}\,sin(\alpha_1/2)\,
 \widetilde{\mathbf{R}}
 (\tilde{\alpha}_1/2, \tilde{\mathbf{n}}_{12})\tilde{\odot}\, 
 \tilde{\mathbf{x}}_1
\end{equation}
where  $\tilde{\odot}$ represents the dual multiplication of a dual 
matrix with a dual vector, $\tilde{\alpha}_1=[\alpha_1,0,0]$, 
$\tilde{\mathbf{n}}_{12}$ the dual version of Eq. (\ref{n12}), which,
in order to be constructed all the involved operations need to be 
promoted to dual operations. Needless to say, such operations are coded in 
the downloadable module. It is interesting to note that neither the sum 
nor the scalar multiplication, need to be dualized, this is, of course 
due to the linearity of the derivative operator.

\subsection{Theoretical calculation of the energy consumption.}
From the theoretical point of view, the energy consumption of the
spherical 4R mechanism can be calculated as follows.

The external forces to the spherical 4R mechanism are,
\begin{itemize}
\item
$\mathbf{R}_1$: The reaction force of the frame on the $\mathbf{x}_1$ 
point.

\item
$\mathbf{R}_2$: The reaction force of the frame on the $\mathbf{x}_4$ 
point.

\item
$\mathbf{F}_g$ : The weight of the mechanism.

\item
$\mathbf{F}_{M}$: The external force due to the motor (the agent who 
generates the rotation of the input link).
\end{itemize}

In our description, the work done by $\mathbf{R}_1$ and $\mathbf{R}_2$ 
is zero, as the points $\mathbf{x}_1$ and $\mathbf{x}_4$ are fixed.
Since $\mathbf{F}_g$ is a conservative force, the work done by this 
force will be  $W_{\mathbf{F}_g} = -\Delta U,$ where, choosing the 
acceleration of gravity as $\mathbf{g} = -g\, \hat{\mathbf{k}}$, the  
gravitational potential energy for the mechanism of mass $m$ is,
\begin{equation}
U(\theta) = m\,g\,z_{cm}(\theta),
\end{equation}
with $z_{cm}(\theta)$ the $z$--coordinate of the center of mass as a 
function of the input angle $\theta$.

From the work-energy theorem, a differential amount of work done by the 
force $\mathbf{F}_{M}$ can be written as
\begin{equation}
\delta W_{\mathbf{F}_{M}}= (dU +  dK) +  \delta E_{\text{others}},
\end{equation}
where $K$ is the kinetic energy and the positive quantity 
$\delta E_{\text{others}}$ represents all the other kinds of energies we 
have not considered, for instance the heat dissipated by friction, etc.

The energy expended by the external agent (the motor) will be

\begin{equation}
\delta E = |(dU +  dK)| + \delta E_{\text{others}}. 
\end{equation} 
Notice that the absolute value is used, this is because the quantity 
$(dU + dK)$ can be negative and it could happen that $\delta 
W_{\mathbf{F}_{M}}$ to be zero\footnote{From a theoretical point of 
view, if the quantity $\delta E_{\text{others}}$ is known, one can 
obtain an angular velocity profile --eigenmotion-- for the input link in 
such a way that $\delta W_{\mathbf{F}_{M}} = 0.$}, which in fact could 
be true. However, this fact does not mean that the motor moves the 
mechanism without consuming energy.  In our case, with a conventional 
motor operating the mechanism, there is no difference between a negative 
work and a positive work since with a conventional motor we cannot reuse 
the energy taken from the mechanism. In the best 
scenario the energy consumption will be $\delta E_{\text{others}}.$

Considering $U$ and $K$ as functions of $\theta$, using a Taylor 
expansion we have
\begin{equation}
U(\theta + d\theta) - U(\theta) =  U'(\theta) d\theta
\end{equation}
 from where 
 \begin{equation}\label{infinidu}
dU =  U'(\theta) d\theta
\end{equation}
and similarly 

\begin{equation}
dK =  K'(\theta) d\theta.
\end{equation}

Thus in a cycle the energy consumption is given by
\begin{equation}\label{Ecoms}
E=\int_0^{2\pi} |U'(\theta) + K'(\theta)|d\theta + E_{\text{others}}.
\end{equation}

Applying this equation to  the balanced ($B$) and unbalanced ($NB$) 
mechanisms, we have,
\begin{equation}\label{DEcoms}
E_{NB}-E_{B} = \int_0^{2\pi} \left(|U'(\theta) + K'(\theta)|_{NB}
- |U'(\theta) + K'(\theta)|_{B} \right)
d\theta  + (E_{\text{others;}NB} -E_{\text{others};B}).
\end{equation} 

A quantification of $ E_{\text{others}}$ is not a trivial task. However
if both mechanisms are constructed as similar as possible (except of 
course by the extensions), the frictional forces are also as reduced as 
possible and if  the deformation energy for both mechanisms are either 
similar or small, then a plausible approximation is that 
$E_{\text{others;}NB} \approx E_{\text{others};B}$. Of course this 
approximation is not in general true, but in our case it can be 
justified\emph{ a posteriori} as our theoretical results are in good 
agreement  with the experimental measure for $E_{NB}-E_{B}$.

\section{Results}\label{Results}

\subsection{Optimum synthesis with minimum acceleration of the center of 
mass}
The multiobjective optimization problem (simultaneous optimization of 
Eqs. (\ref{objectivef}) and (\ref{fcm})) was solved by using the 
weighted sum method and the evolutionary algorithm known as Differential 
Evolution \cite{Price_Storn2005}. Specifically, we use the DE/rand/1/bin 
strategy with the dither variant. For the numerical results we have used 
a population of 50 individuals and a number of generations of 10\,000 and 
a crossover probability of 0.9.
  
Assuming a constant angular velocity of 1 rad/s for the input link (we 
use a PID controller to regulate this velocity), the DE method gives as 
result the following values: 
$ \mathbf{x}_1 = [ 1.0000 , 0.0000, 0.0000]$, 
$ \mathbf{x}_4 = [0.54462, 0.80817, 0.22413]$,
$\alpha_1=0.40144$ rad,
$\alpha_2=0.82035$ rad,
$\alpha_3=0.92505$ rad, 
$ \beta = 0.23066$ rad,
$ \gamma = 0.47437$ rad,
$ e_1  = 4.46818$ rad,
$ e_2  = 0.94312$ rad,
$ e_3  = 0.24879$ rad and
$ e_4  = 4.30884$ rad.
\begin{table}[htb]
\caption{\label{Tpar} Parameters of the constructed spherical 4R 
mechanism.}
\vspace*{0.1cm}
\centering
\scalebox{0.9}{
\begin{tabular}{cccccccccccccccc}
\toprule 
 \multicolumn{3}{c}{$\mathbf{x}_1$}&&\multicolumn{3}{c}{$\mathbf{x}_4$}  &$\alpha_1$
  & $\alpha_2$ &$\alpha_3$ & $\beta$ & $\gamma$ & $e_1$ & $e_2$ & $e_3$ & $e_4$\\
 \cline{1-3} \cline{5-7} 
  1.00 &0.00 & 0.00   &&  0.39 &0.92&0.00 &0.40 & 0.88 &1.39&0.34 &0.40 &1.92 & 1.06 & 0.00 & 2.79\\
\bottomrule 
\end{tabular} 
}
\end{table} 
\begin{figure}[htb]
\begin{center}
\includegraphics[scale=0.2]{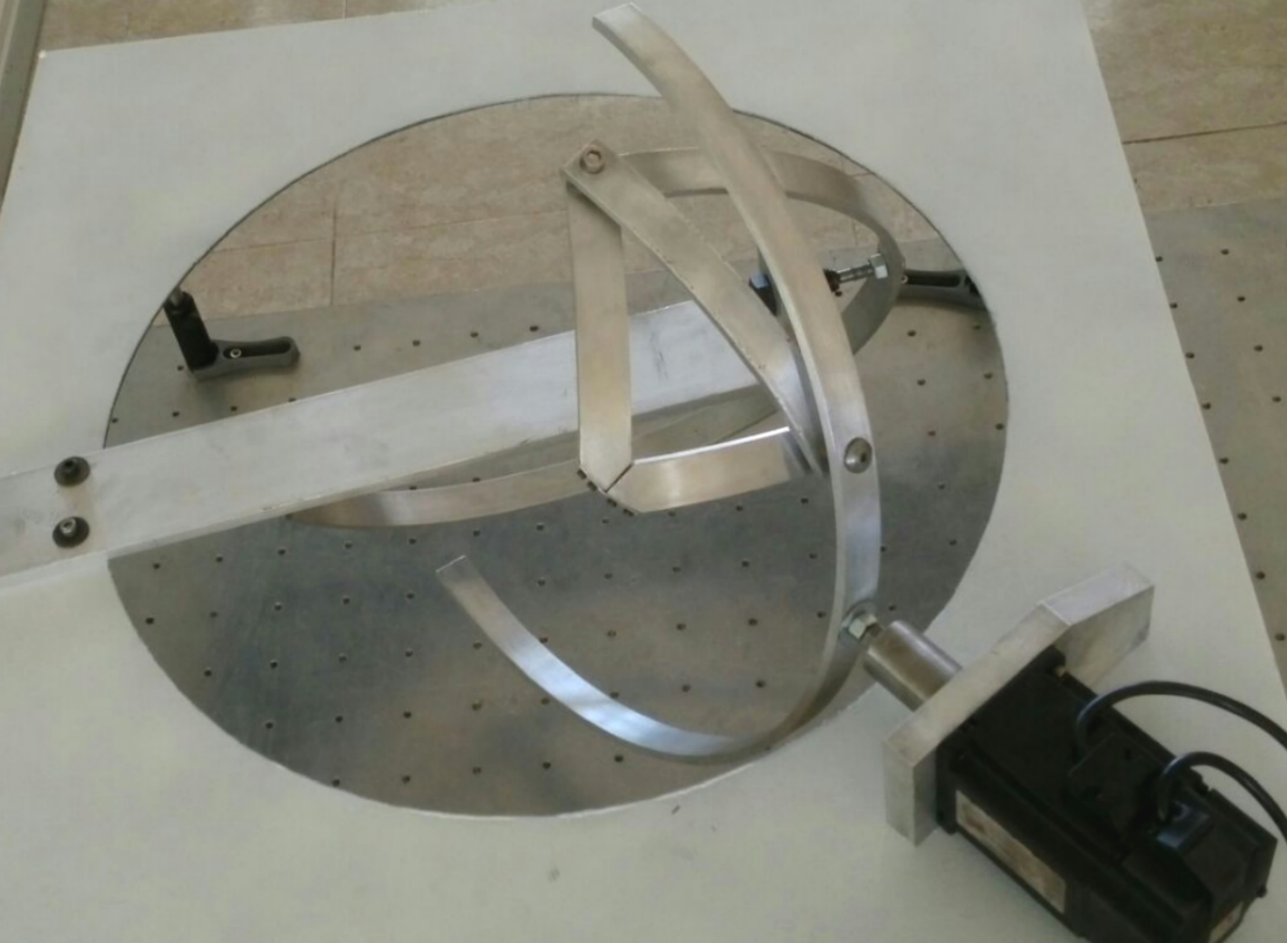} 
\caption{Constructed spherical 4R mechanism. }
\label{figure3}
\end{center}
\end{figure}
\begin{figure}[htb]
\begin{center}
\includegraphics[scale=0.8]{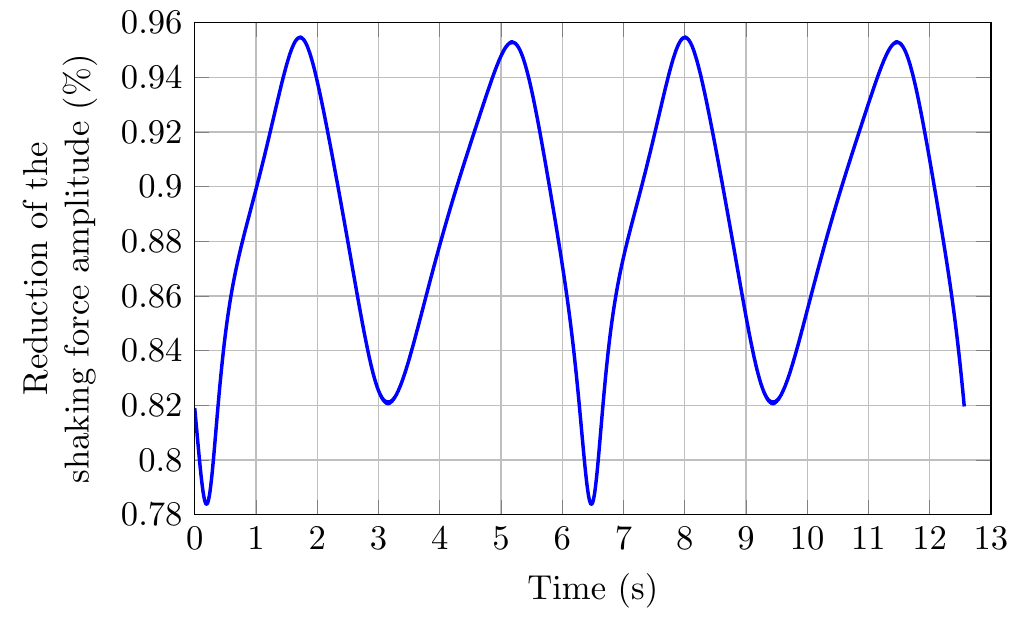} 
\caption{Reduction of the shaking force amplitude for the mechanism with 
minimum center of mass acceleration. The mean value is of 89\%. }
\label{figure4}
\end{center}
\end{figure}
With these design parameters we obtain $f_{path}=0.00002$ m and 
$f_{cm}=0.01351$ m/s$^2$.  Unfortunately it is not possible to build 
this mechanism, as the links (with extensions) would collide with the 
supports for the frame points ($\mathbf{x}_1$ and $\mathbf{x}_4$). A 
feasible mechanism can be synthesized demanding that the $z$-coordinate 
for the points  $\mathbf{x}_1$ and $\mathbf{x}_4$ to be zero, and 
choosing horizontal supports for the frame points, thus avoiding 
collisions with the extensions. With these considerations, we  obtain 
the mechanism whose parameters are shown in Table \ref{Tpar}. 
Fig. \ref{figure3} shows a first prototype of the constructed 
mechanism. For this mechanism  $f_{path}=0.00053$ m, while 
$f_{cm}=0.014$ m/s$^2$. It is worthwhile to note that if no extensions 
are considered for the mechanism; i.e. $e_i=0$ for $i=1,...,4$; the 
multiobjective optimization method yields $f_{cm}=0.28$  m/s$^2$ which 
is twenty times bigger than the corresponding value for
the mechanism with extensions. As a result, the shaking forces are 
minimized. From Fig. \ref{figure4} we can see that the reduction of the 
shaking force amplitude for the balanced mechanism, reaches values 
of $ 95\%$ and in the worst scenario of $ 78\%$, having a mean value of 
$89\%$.  Notice that this reduction is 
independent (from a theoretical point of view) of the used angular 
velocity for the input link ($\dot{\theta}$). Because if $\dot{\theta}$ 
in Eq. (\ref{rcmdt2}) is constant, the ratio between ACoM for the balanced 
and unbalanced mechanisms, does only depend on the ratio of
their second partial derivatives with respect to $\theta$. 

\subsection{Experimental validation}
\begin{figure}[htb]
\begin{center}
\includegraphics[scale=0.8]{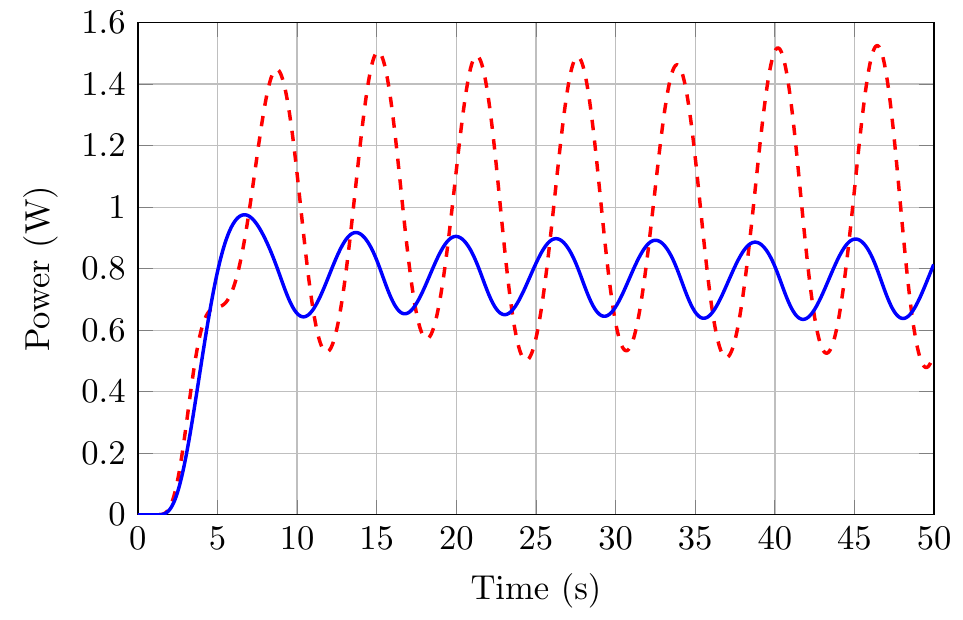} 
\caption{Experimental power consumption comparison between the ACoM 
balanced (solid line) and unbalanced (dashed line) mechanisms.}
\label{figure5}
\end{center}
\end{figure}
\begin{figure}[htb]
\begin{center}
\includegraphics[scale=0.8]{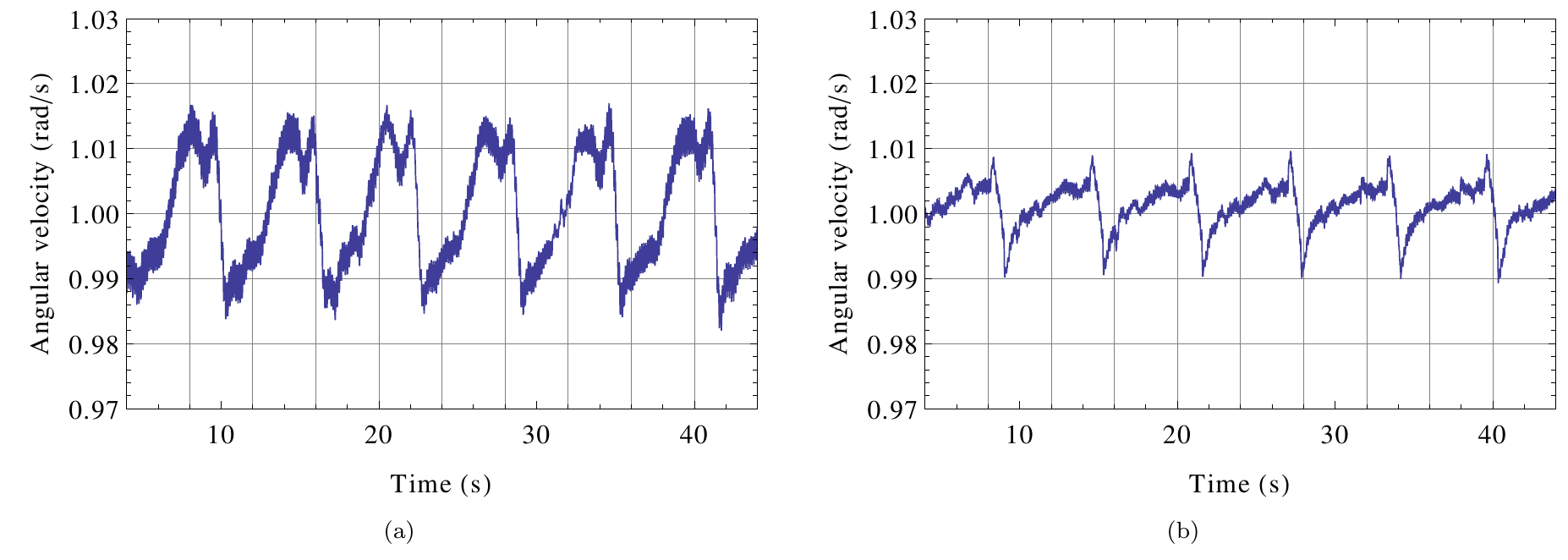} 
\caption{Controlled velocity for the unbalanced mechanism (a) 
and the ACoM balanced mechanism (b).}
\label{figure6}
\end{center}
\end{figure}
In order to validate our theoretical results (shown in Fig 
\ref{figure4}), either a measurement of the ACoM or of the external 
forces is required. Such experimental measurements are not easy to be 
performed. Fortunately an experimental validation of theoretical results 
can be made by measuring the difference in the energy consumption  
$E_{NB}-E_{B}$, of the balanced and unbalanced mechanisms. The fact 
that the quantity under the integral sign in Eq. (\ref{DEcoms}) depends 
on the velocity and the acceleration of the center of mass allows to 
compare such theoretical quantity with its experimental counterpart.
To this end, we have manufactured 
both mechanisms with and without extensions, considering a sphere of 
mean radius\footnote{The links of the spherical 4R mechanism where 
constructed in concentric spheres of different radius. The deviation in 
the radius for the links depends on the thickness of the links, in our 
case 6.35 mm.} equal to 0.23 m. Specifically, the radii for the input link 
the coupler link and the output links were of 0.250 m, 0.235 m and 0.210 m 
respectively. The radius for the links forming the triangle (Fig. \ref{figure3}) 
for the coupler point was of 0.255 m.
The links were constructed with aluminum 
slab of 6.35 mm thickness and 2.54 cm width. For both mechanisms a dc 
motor is used as actuator to the input link, which is controlled by a 
PID law with a reference of 1 rad/s. The energy required to operate the 
mechanism is considered equal to the electrical energy required to 
actuate the motor, which was obtained from direct measurement of voltage 
and current at its terminals. In Fig \ref{figure5}, we show the power 
consumption for both mechanisms. Fig \ref{figure6}, presents the time 
behavior of the angular velocity of the input link of both mechanisms 
(which is obtained as the output of the filter $s/(1\,+\,0.01\, s)$, 
whose input comes from a digital encoder attached to the joint of the 
input link), we can observe that the regulation objective is not 
absolutely reached, however the oscillations around the reference are 
smaller for the case of the balanced mechanism. 

After an integration of the experimental data showed in Fig. 
\ref{figure5}, we have obtained a deviation of only $5\%$ with respect
to a theoretical calculation of $E_{NB}-E_{B}$ using Eq. (\ref{DEcoms})
and neglecting $E_{\text{others;}NB} -E_{\text{others};B}$.

\section{Discussion}\label{Discussion}
Before concluding, some comments on the used derivative method and the
energy consumption are worthwhile. Below we present a brief discussion.

\subsection*{Numerical derivatives: Dual numbers vs Finite Differences}
Using finite differences (FD), the second derivative of a function can 
be calculated as,
\begin{equation}\label{diff1}
f''(x) = \frac{f(x+h)-2\,f(x)+f(x-h)}{h^2}+O(h^2).
\end{equation}
In order to obtain the second derivative using this expression, three 
evaluations of the function $f$ are required. Since in the dual number 
approach we only need one evaluation, it is reasonable to think  that 
this method will be more efficient. However, we cannot say that the dual 
number approach will be three times faster than the finite difference 
method, as the involved evaluations do not correspond to the same 
functions.  In our case, for 5000 points equally spaced in $(0,2\pi)$, 
the matrix $[ \mathbf{\dot{r}}_{cm},~ \mathbf{\ddot{r}}_{cm}]$ was 
calculated 2.5 times faster by using the dual numbers approach instead 
of the FD method. 

Now, from Eq. (\ref{diff1}) a truncation error of order $h^2$ is 
present. One could think that this problem can be solved by taking a 
small value for $h$, but then subtractive cancellation errors will 
appear. For example taking $h=1\times 10^{-6}$ and using double 
precision for the variables, the error in the calculation of the ACoM is 
not of the order $h^2$ but of the order of $10^{-2}$. This situation can 
be alleviated by writing a more elaborated formula for $f''$ but there 
is a cost to pay, we will need more evaluations of the function $f$. 
These problems are not present when the dual numbers approach is used to 
calculate the derivatives. Moreover, this fact is not the only advantage 
of the dual number approach. For instance, one could be interested in 
obtaining the derivative of a function that is the solution of some 
equation which can only be known numerically. In this case, we can 
implement the numerical solution method in the context of the dual 
number approach and then both, the solution and its derivative will be 
directly obtained.

\subsection*{Energy consumption}
Even when minimization of ACoM does not imply a constant gravitational 
potential energy --excepting if ACoM is zero and 
for a mechanism operating in cycles\footnote{Let us assume that ACoM is 
zero, this means that the velocity of the center of mass is a constant. 
Then, in a cycle the mechanism must return to its initial point, 
therefore the only  allowed constant for the velocity of the center of 
mass is zero. Thus GPE is constant. Notice that the converse is not 
true. }--, we have found (integrating the absolute value of 
Eq. (\ref{infinidu})) that the variations in the potential energy are 
reduced about $92\%$ for the balanced mechanism.

It is worthwhile to mention that not only the shacking forces were 
reduced by balancing the mechanism but also the energy consumption 
(obtained by integrating the experimental data of Fig. \ref{figure5}) 
was reduced by a 21 \%. Of course this is not a rule since we are not 
optimizing Eq. (\ref{Ecoms}). Moreover, unlike the percentile reduction 
of the total external force, the percentile reduction of the energy 
consumption depends on the angular velocity of the input link.

\section{Conclusions}\label{conclu}
By implementing the center of mass vector in its dual version, a simple 
yet effective method for balancing a spherical 4R mechanism 
using the counterweights method has been presented. The minimization of 
the magnitude of the center of mass acceleration vector has been 
conducted with the Differential Evolution algorithm.  The synthesis 
process for the prescribed path generation task turn out to be highly 
efficient giving excellent results. The reduction of the force shaking 
amplitude, has reached a mean value of 89\% while the variations in the 
gravitational potential energy have been reduced about 92\%. The 
synthesized mechanisms, the balanced and its unbalanced 
counterpart, were constructed in order to experimentally measure the 
power consumption and hence the energy needed for both mechanisms to 
perform the requested path. By doing this, an experimental validation of 
theoretical results has been conducted. As a side effect, it is 
interesting to note that the energy used to operate the  balanced 
mechanism was reduced by 21\% with respect to its unbalanced 
counterpart. Nevertheless this cannot be considered as a rule as a 
minimization of ACoM does not necessarily implies a minimization of 
Eq (\ref{Ecoms}).

\section*{Acknowledgments}
Authors thank financial support from the mexican PROMEP and the National 
System for Researchers. Authors also acknowledge the anonymous reviewers 
whose comments and suggestions greatly improved this work.

\end{document}